\documentclass[preprint,amsmath,amssymb,prb,floatfix,superscriptaddress]{revtex4}

\usepackage[dvips]{graphicx}
\usepackage{dcolumn}
\usepackage{bm}
\usepackage{mathrsfs}

\begin{document}
\title{Hybridization and superconducting gaps in heavy-fermion superconductor PuCoGa$_5$ probed via the dynamics of photoinduced quasiparticles}
\author{D. Talbayev}
 \email{diyar.talbayev@yale.edu}
 \affiliation{Los Alamos National Laboratory, Los 
Alamos, NM 87545, USA}
\author{K.S. Burch}
 \affiliation{Department of Physics and Institute of Optical Sciences, University of Toronto, 60 St. George Street, Toronto, ON M5S 1A7, Canada}
\author{Elbert E.M. Chia}
 \affiliation{Division of Physics and Applied Physics, School of Physical and Mathematical Sciences, Nanyang Technological University, Singapore 637371, Singapore}
\author{S.A. Trugman}
 \affiliation{Los Alamos National Laboratory, Los Alamos, NM 87545, USA}
\author{J.-X. Zhu}
 \affiliation{Los Alamos National Laboratory, Los Alamos, NM 87545, USA}
\author{E.D. Bauer}
 \affiliation{Los Alamos National Laboratory, Los Alamos, NM 87545, USA}
\author{J. A. Kennison}
 \affiliation{Los Alamos National Laboratory, Los Alamos, NM 87545, USA}
\author{J. N. Mitchell}
 \affiliation{Los Alamos National Laboratory, Los Alamos, NM 87545, USA}
\author{J. D. Thompson}
 \affiliation{Los Alamos National Laboratory, Los Alamos, NM 87545, USA}
\author{J.L. Sarrao}
 \affiliation{Los Alamos National Laboratory, Los Alamos, NM 87545, USA}
\author{A.J. Taylor}
 \affiliation{Los Alamos National Laboratory, Los Alamos, NM 87545, USA}
\date{\today}

\newcommand{\cm}{\:\mathrm{cm}^{-1}}
\newcommand{\T}{\:\mathrm{T}}
\newcommand{\mc}{\:\mu\mathrm{m}}
\newcommand{\ve}{\varepsilon}
\newcommand{\dg}{^\mathtt{o}}
\newcommand{\eph}{$e$-$ph\ $}

\begin{abstract}
We have examined the relaxation of photoinduced quasiparticles in the heavy-fermion superconductor PuCoGa$_5$. The deduced electron-phonon coupling constant is incompatible with the measured superconducting transition temperature $T_c=18.5$ K, which speaks against phonon-mediated superconducting pairing. Upon lowering the temperature, we observe an order-of-magnitude increase of the quasiparticle relaxation time in agreement with the phonon bottleneck scenario - evidence for the presence of a hybridization gap in the electronic density of states. The modification of photoinduced reflectance in the superconducting state is consistent with the heavy character of the quasiparticles that participate in Cooper pairing.
\end{abstract}

\maketitle

The discovery of relatively high-temperature superconductivity (SC) in the Pu-based compounds PuCoGa$_5$ ($T_c$=18.5 K) and PuRhGa$_5$ ($T_c$=8.7 K) has renewed interest in actinide materials research\cite{sarrao:297, wastin:2279}. The Pu-based superconductors crystallize in the same HoCoGa$_5$-type tetragonal lattice structure as the Ce-based series of compounds (CeRhIn$_5$, CeCoIn$_5$, and CeIrIn$_5$) commonly referred to as ``115" materials. The Ce-based 115 compounds, CeIrIn$_5$ ($T_c$=0.4 K) and CeCoIn$_5$ ($T_c$=2.3 K), display superconductivity at ambient pressure\cite{petrovic:354, petrovic:337, sarrao:51013}. Both Ce- and Pu-based 115 compounds display the heavy-fermion behavior resulting from the influence of 4$f$ (Ce) and 5$f$ (Pu) electrons. The most intriguing question concerns the origin of SC in the 115 materials. In the Ce series, the $d$-wave symmetry of the SC order parameter and the proximity of SC order to magnetism have led to a widespread belief that the unconventional SC is mediated by antiferromagnetic spin fluctuations\cite{sarrao:51013}. In the Pu compounds, two possible scenarios regarding the SC mechanism have been considered: one approach favors a magnetically mediated unconventional SC similar to that in CeCoIn$_5$\cite{curro:622}. In the other scenario, the conventional SC is mediated by phonons\cite{bang:104512, jutier:24521}, where the strength of the electron-phonon ($e$-$ph$) coupling $\lambda$ is the crucial parameter that sets the SC transition temperature $T_c$. 

In this Letter, we present a measurement of the \eph coupling constant $\lambda$ via the pump-probe optical study of photoinduced quasiparticle relaxation time. We find that \eph coupling ($\lambda=0.2-0.26$) is too weak to explain the high $T_c$ of PuCoGa$_5$ and that phonon-mediated superconductivity is unlikely in this material. Upon lowering the temperature in the normal state ($T>T_c$), we find an order-of-magnitude increase in the relaxation time consistent with a phonon bottleneck, similar to other heavy-fermion materials\cite{demsar:281}, which provides the first optical evidence for a hybridization gap in the electronic density of states (DOS) of PuCoGa$_5$. Below $T_c$, a SC gap opens at the Fermi level, and the resulting quasiparticle dynamics confirms that the same heavy quasiparticles detected in the normal state also participate in the SC pairing. Our results are consistent with theoretical investigations\cite{opahle:157001, maehira:207007}, which find that the electronic structure is dominated by cylindrical sheets of Fermi surfaces with large 5$f$ electron character, suggesting that the delocalized 5$f$ electrons of Pu are responsible for an unconventional (nonphonon) mechanism of superconductivity\cite{monthoux:1177}.

\begin{figure}[ht]
\begin{center}
\includegraphics[width=5.0in]{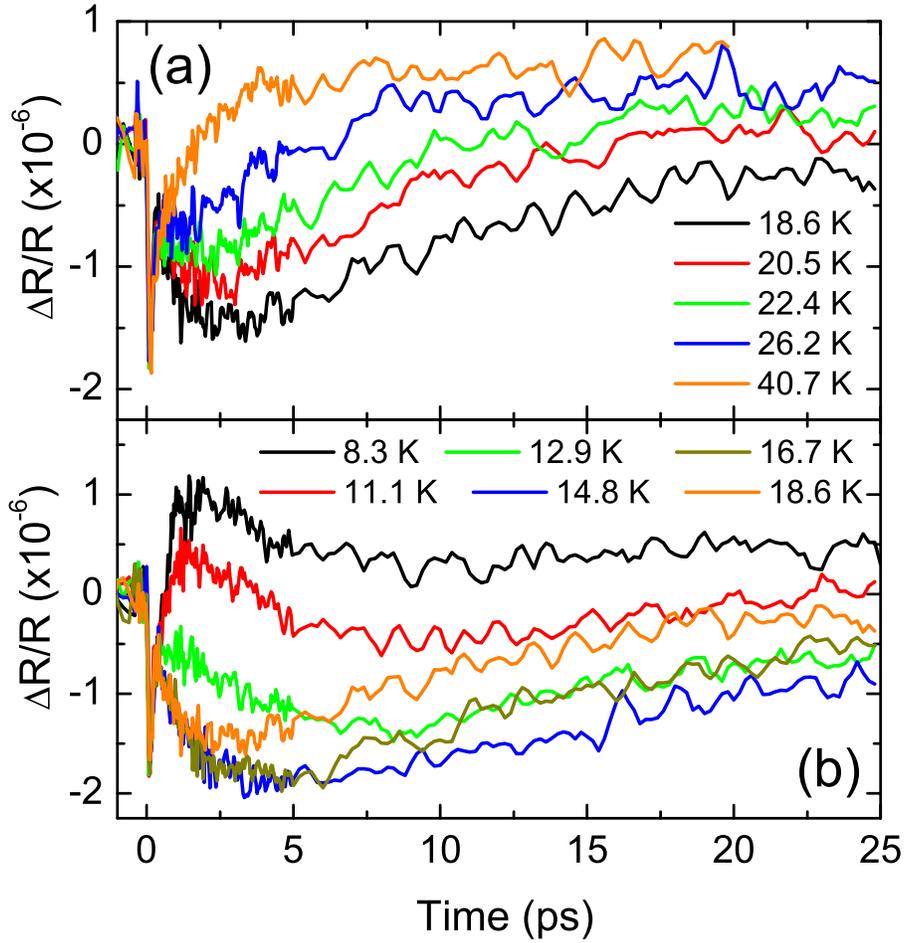}
\caption{\label{fig:ppspectra}(Color online) Photoinduced reflectance of PuCoGa$_5$ at various temperatures. The data were smoothed using 3-point adjacent averaging to reduce the visual overlap due to noise between curves. As-collected data can be found in Fig.~\ref{fig:fit}.}
\end{center}
\end{figure}

In our experiment, 45-femtosecond optical pulses are used to photoexcite (pump) the material. The material's response is monitored using a weaker optical pulse (probe) that allows the measurement of pump-induced changes in the material's reflectance (photoinduced reflectance). By varying the time delay between the pump and probe pulses, we measure the evolution of the photoinduced reflectance with time resolution $\leq100$ fs (Fig.~\ref{fig:ppspectra}). Our pump-probe study was conducted on a clean as-grown surface of PuCoGa$_5$ single crystal grown in Ga flux\cite{sarrao:297}.  We used an 80-MHz repetition rate Ti:sapphire laser producing 800-nm (1.55 eV) light pulses that were split into pump and probe parts. The probe pulses were focused onto a 30 $\mu$m spot, while the size of the pump spot was 60 $\mu$m. We used average pump powers of 0.5 and 1 mW, corresponding to pump fluences of 0.015 and 0.03 $\mu$J/cm$^2$. The probe average power was at least 10 times lower. The small spot sizes and the high repetition rate provide a resolution of at least 1 part in $10^6$ while minimizing sample heating by the pump.

In a metal, the pump pulse creates an initial nonthermal distribution of electrons due to electronic absorption of photons\cite{groeneveld:11433,averitt:1357}. The subsequent relaxation of this system is governed by electron-electron ($e$-$e$) and \eph collisions. At high temperatures ($T \geq \theta_D$/5) in metals\cite{ groeneveld:11433}, the $e$-$e$ collisions happen at a much faster rate than the $e$-$ph$ collisions. The nonthermal distribution thermalizes through the $e$-$e$ collisions before any significant exchange of energy between the electrons and the lattice can take place, which leads to different effective electronic and lattice temperatures, $T_E$ and $T_L$. The much slower equilibration of $T_E$ and $T_L$ proceeds with a characteristic time scale $\tau_{EL}$ set by the strength of the $e$-$ph$ coupling. The relaxation time of photoinduced reflectance corresponds to the \eph relaxation time $\tau_{EL}$. Allen\cite{allen:1460} showed that $\tau_{EL}$ is related to the \eph coupling constant $\lambda$ as
\begin{equation}
\tau_{EL}=\pi k_B T_E/3\hbar \lambda\left\langle \omega^2\right\rangle ,
\label{eq:allen}
\end{equation}
where $\left\langle \omega^2\right\rangle$ is the second moment of the phonon spectrum. In PuCoGa$_5$ ($\theta_D=240$ K\cite{sarrao:297}), we measure the room-temperature relaxation time $\tau_3$ by fitting a single-exponential function $y=A_3\exp(-t/\tau_3)$ to the recorded spectra and find $\tau_3=\tau_{EL}=0.64\pm 0.01$ ps.

The coupling constant $\lambda$ is calculated using Eq. (\ref{eq:allen}) and $T_E=300$ K, since the initial pump-induced rise in electronic temperature $\Delta T_E(0)=T_E(0)-T_L(0)\ll T_L$. In the absence of detailed knowledge about which phonon mode might be "responsible" for superconductivity in this material, we assume $\left\langle \omega \right\rangle^2 \leq \left\langle \omega^2 \right\rangle \leq \theta_D^2$, where $\left\langle \omega \right\rangle = 212$ K was calculated by Piekarz $et$ $al.$\cite{piekarz:14521}, and determine $0.2\leq \lambda \leq 0.26$. Our estimate of $\lambda$ represents the first experimental determination of the \eph coupling in this material and can be compared to $\lambda=0.7$ found in the $ab$ $initio$ calculation of Piekarz $et$ $al.$  By assuming a strong coupling limit\cite{curro:622} and using the simplified Allen-Dynes\cite{allen:905} formula for the SC transition temperature
\begin{equation}
T_c \approx \theta_D \exp\left[ - \frac{1.04(1+\lambda)}{\lambda-\mu^*(1+0.62\lambda)}\right]
\label{eq:mcmillan}
\end{equation}
with $\lambda=0.26$ and $\mu^*=0.1$ as a representative value of the effective Coulomb repulsion, we find $T_c \approx 0.027$ K ($\approx 1.55$ K in the upper limit of $\mu^*=0$). An estimate using Rowell's formula\cite{rowell:1131} linear in $\lambda$ gives $T_c=\theta_D(\lambda-0.25)/20=0.12$ K. These estimates of $T_c$ differ from the measured $T_c=18.5$ K by nearly 3 orders of magnitude and reinforce the conclusion from similar estimates by Piekarz $et$ $al.$\cite{piekarz:14521} and Bang $et$ $al.$\cite{bang:104512} (who used $\lambda=1$) that a phonon origin of SC pairing in PuCoGa$_5$ is highly unlikely. 

\begin{figure}[ht]
\begin{center}
\includegraphics[width=5.0in]{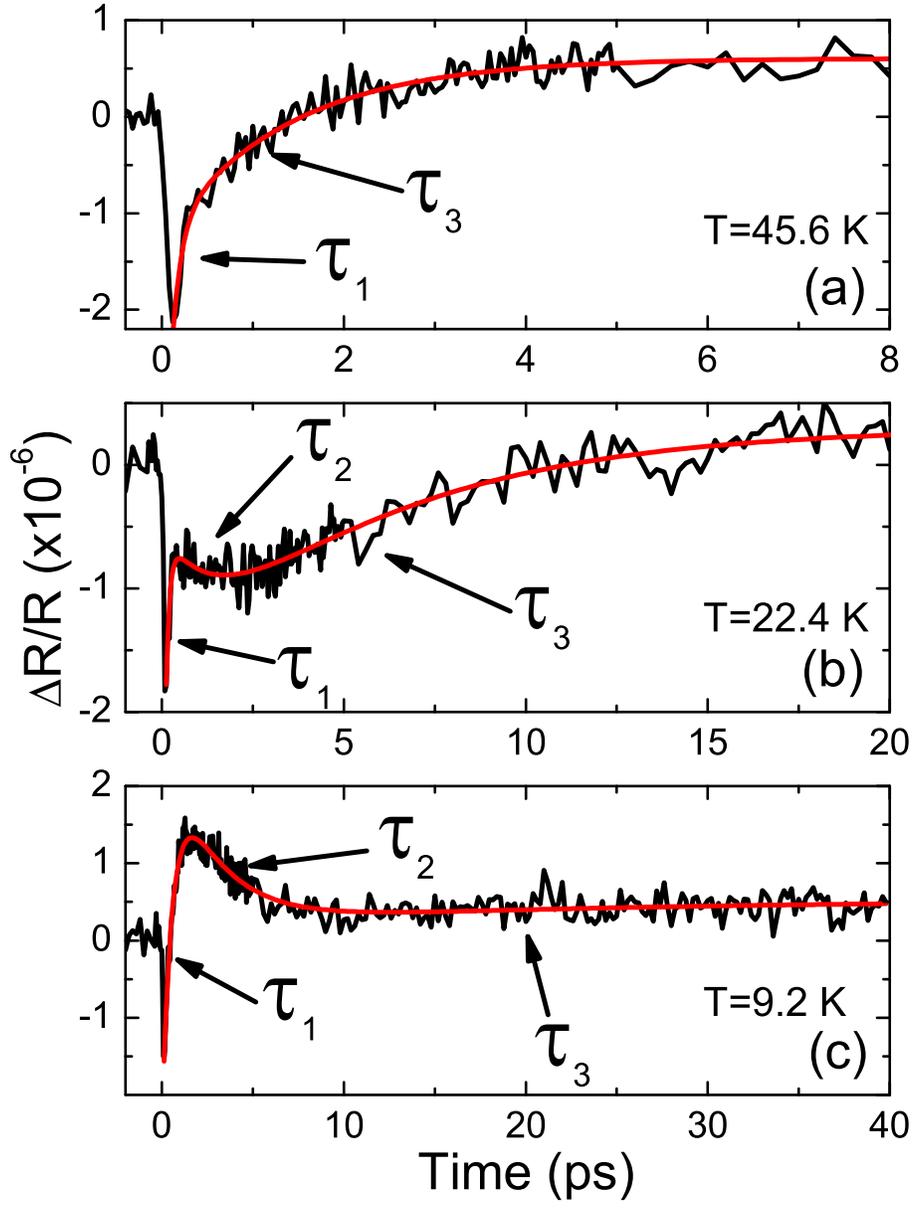}
\caption{\label{fig:fit}(Color online) (a) Photoinduced reflectance in the normal state  at $T>25$ K. Smooth line is a 2-exponential fit of the measured reflectance. $\tau_1$ denotes the initial dynamics due to quasiparticle thermalization, $\tau_3$ denotes \eph relaxation. (b) Photoinduced reflectance at $18<T<25$ K. Smooth line is a 3-exponential fit of the measured reflectance. $\tau_2$ indicates the initial dynamics due to the emission of excess high-frequency phonons. (c) Photoinduced reflectance at $T<T_c\approx18$ K. The amplitude $A_3$ of the $\tau_3$ process is greatly suppressed. Smooth line is a 3-exponential fit.}
\end{center}
\end{figure}

In conventional SCs, the room-temperature pump-probe measurements of $\lambda$ were in excellent agreement with $\lambda$ deduced from their SC transition temperatures\cite{brorson:2172}. Several pump-probe studies of SC cuprate Bi$_2$Sr$_2$CaCu$_2$O$_{8+\delta}$ measured the \eph relaxation time and found that the corresponding $\lambda$  cannot account for the high $T_c$ of that material ($\lambda<0.25$ due to Perfetti $et$ $al.$\cite{perfetti:197001}, $\lambda<0.17$ due to Zhu $et$ $al.$\cite{zhu:unpublished}). The experimentally determined $\lambda$ in Bi$_2$Sr$_2$CaCu$_2$O$_{8+\delta}$ compares favorably with the theoretical value calculated\cite{bohnen:104} for YBa$_2$Cu$_3$O$_7$ ($\lambda=0.27$), which is expected to exhibit a similar strength of the \eph coupling. 

\begin{figure}[ht]
\begin{center}
\includegraphics[width=5.0in]{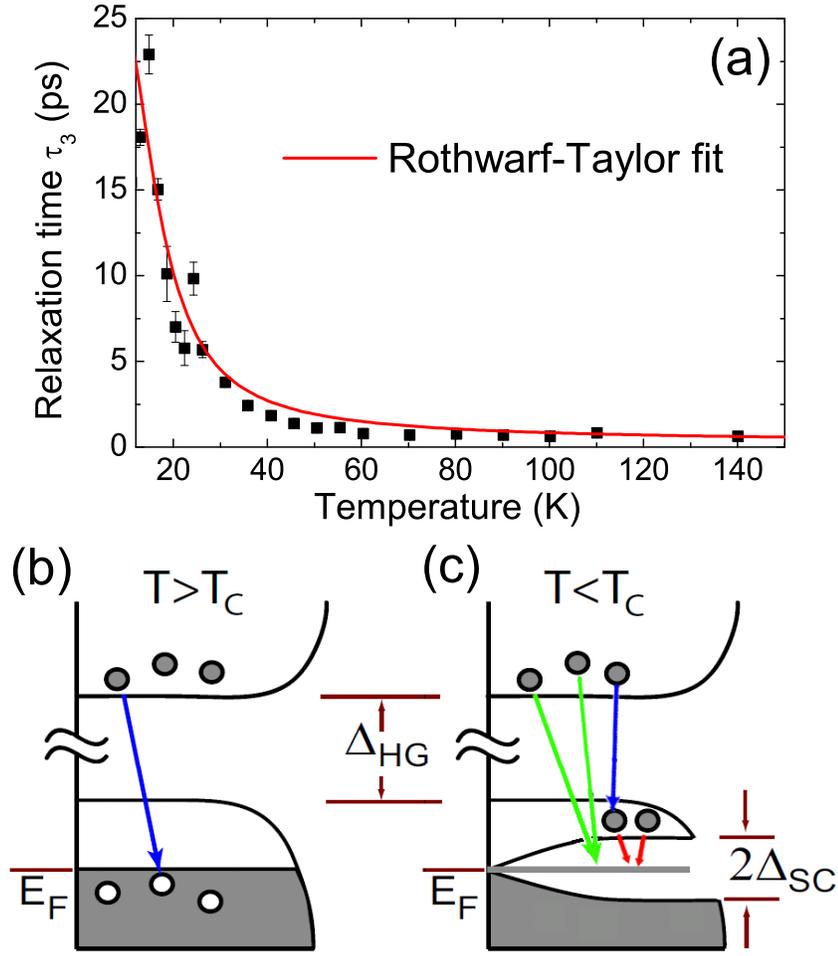}
\caption{\label{fig:elecdiag} (Color online) (a) Temperature dependence of relaxation time $\tau_3$. (b),(c)  A diagram of an electronic structure that is consistent with our experiments. (b) Electron-hole pair recombination across the hybridization gap above $T_c$. Rothwarf-Taylor bottleneck occurs. (c) Quasiparticle relaxation in the superconducting state ($T < T_c$) occurs via multiple channels.}
\end{center}
\end{figure}

Upon lowering the temperature, the relaxation time in PuCoGa$_5$ starts to gradually increase at $\simeq100$ K, and a very fast initial relaxation process appears at $\simeq55$ K (Fig~\ref{fig:fit}(a)). We extract the associated relaxation time by fitting the measured reflectance with a 2-exponential decay $y=A_1\exp(-t/\tau_1) + A_3\exp(-t/\tau_3)$, where the first term describes the initial fast relaxation and the second term corresponds to the \eph relaxation at higher temperatures. We find $\tau_1=0.10\pm0.02$ ps at 45 K, the value that is very weakly dependent on temperature. We attribute this process to a thermalization of initially excited hot electrons through $e$-$e$ collisions due its short characteristic time. The relaxation time $\tau_3$ increases by an order of magnitude between 100 and 20 K (Figs.~\ref{fig:ppspectra}(a) and ~\ref{fig:elecdiag}(a),(b)). Similar increase has been found in other heavy-fermion metals\cite{demsar:027401, demsar:037401, demsar:281, burch:26409} and was ascribed to the hybridization gap, $\Delta_{HG}$, in the electronic DOS, which results from the hybridization between the conduction band and the localized $f$-electron states\cite{demsar:281}. After the initial photoexcitation, the $e$-$e$ collisions lead to a quasiparticle multiplication effect\cite{kabanov:1497, demsar:281}: each photon of energy $E_0$ creates $E_0/\Delta_{HG}$ electron-hole ($e$-$h$) pairs. The subsequent decay of the $e$-$h$ pair population is responsible for long relaxation of photoinduced reflectance (blue arrow in Fig.~\ref {fig:elecdiag}(b)).

The time evolution of the $e$-$h$ pair and the high-frequency phonon (HFP) densities has been modeled using a system of coupled kinetic equations, known as the Rothwarf-Taylor (RT) model\cite{rothwarf:27, kabanov:147002} and originally used to describe quasiparticle relaxation in SCs. The HFPs are intimately involved in the decay of $e$-$h$ pairs, as a phonon with energy $\omega_{ph}\geq\Delta_{HG}$ is created upon $e$-$h$ recombination. Since HFPs can subsequently excite $e$-$h$ pairs, the decay of the $e$-$h$ pair population is governed by the decay of the phonons. This process is often referred to as a phonon bottleneck. The RT model provides a means to calculate the temperature dependence of the relaxation time of photoinduced reflectance, which is governed by the thermally excited density of $e$-$h$ pairs and HFPs, the bare rate of $e$-$h$ pair recombination and creation, and the anharmonic decay time of the HFPs. The solution of the RT model is\cite{kabanov:147002}
\begin{equation}
\tau(T)=\tau_0\left\{\delta[\epsilon n_T+1]^{-1}+2n_T\right\}^{-1},
\label{eq:rt}
\end{equation}
where $n_T=\sqrt{T}\exp(-\Delta_{HG}/T)$ is a thermal quasiparticle density, and $\tau_0$, $\delta$, $\epsilon$, and $\Delta_{HG}$ are temperature independent fitting parameters. Figure~\ref{fig:elecdiag}(a) shows a RT fit of the temperature dependence of the $e$-$h$ recombination time $\tau_3$ where the magnitude of the hybridization gap is found to be $\Delta_{HG}= 44\pm7$ K ($\approx31\cm$). Since $\tau_3$ is determined by the indirect gap, the $\Delta_{HG}$ values deduced from pump-probe measurements in heavy-fermion metals tend to be smaller than those found in optical conductivity studies\cite{demsar:037401, demsar:281, burch:26409} that probe the direct gap. The temperature dependence of the relaxation time $\tau_3$ of Fig.~\ref{fig:elecdiag}(a) is a signature of a gap in the electronic DOS, thus providing the first spectroscopic evidence of hybridization gap in PuCoGa$_5$.

At temperatures $18<T<25$ K, another initial process can be distinguished before the $e$-$h$ recombination takes place (Fig.~\ref{fig:fit}(b)). This second process, after the initial fast recovery, further lowers the reflectance before the $e$-$h$ recombination starts (at the delay of ~2-3 ps in Fig.~\ref{fig:fit}(b)). We extract the characteristic time $\tau_2$ of the second process by fitting the data at these temperatures with a 3-exponential decay $y=A_1\exp(-t/\tau_1) + A_2\exp(-t/\tau_2) + A_3\exp(-t/\tau_3)$, where the first and last terms correspond to the previously described quasiparticle thermalization and $e$-$h$ pair recombination. Thus extracted $\tau_2$ is in the 1-4 ps range. Similar long (up to 10 ps) rise-time dynamics was observed below 25 K in several other heavy-fermion metals\cite{demsar:281}, including CeCoIn$_5$, and the superconductor MgB$_2$\cite{demsar:267002}. In those studies, the long rise time was attributed to the competition between the $e$-$ph$ and $e$-$e$ scattering channels at low temperatures. The same scenario may be realized in our measurements: during the initial dynamics both $e$-$e$ and \eph scattering take place, leading to HFP density that exceeds the thermal phonon density. These excess HFPs then decay through quasiparticle excitation (\eph channel), thus effectively slowing down quasiparticle multiplication.

We now consider the dramatic changes in the photoinduced reflectance that happen below the SC transition temperature $T_c\approx 18$ K. These are best described by the reduction of the amplitude $A_3$ of the $e$-$h$ pair recombination signature. In the low-temperature limit of the RT model and at a fixed pump pulse energy, the amplitude $A_3$ is determined by the number of photoexcited quasiparticles, i.e., the amplitude is expected to plateau as the temperature is lowered\cite{kabanov:147002}. A similar behavior is expected of the relaxation time $\tau_3$. However, this scenario contradicts the measured reflectance spectra. According to the RT model, at $T<10$ K, we expect the $\tau_3 \geq 22.9$ ps and $A_3 \geq 2.5$x$10^{-6}$, where 22.9 ps and $2.5$x$10^{-6}$ are the values measured at 14.8 K. Figure~\ref{fig:fit}(c) shows a fit of the 9.2-K spectrum using $\tau_3 =22.9$ ps as a fixed parameter, which gives $A_3(9.2$ K$)=-0.33$x$10^{-6}$, much lower than $A_3$ at $14.8$ K. The reduced $A_3$ signifies a modification of the RT response due to the opening of the SC gap in the electronic spectrum. Such modification is most readily explained if the superconductivity develops on the same heavy electron band that is responsible for the RT response above $T_c$. Therefore, the modified RT response below $T_c$ suggests that the quasiparticles that form the SC condensate are the same heavy quasiparticles formed by the opening of the hybridization gap. This conclusion agrees with the measurements of the specific heat and the critical field $H_{c2}$ that indicate the enhanced effective mass of the condensing quasiparticles\cite{sarrao:297, bauer:147005}. Thus, our observations provide the spectroscopic evidence for superconductivity in PuCoGa$_5$ developing on the heavy quasiparticle band.

To illustrate our finding, we assume that the Fermi level is located below the hybridization gap in the picture of a single conduction band hybridized with a single Pu 5$f$ band. (Alternatively, one could place the Fermi level above the hybridization gap and construct an equally valid description where electrons and holes are interchanged.) Below $T_c$, a SC gap appears at the Fermi level (Fig.~\ref{fig:elecdiag}(c)). The photoexcitation of a conventional SC creates quasiparticles that accumulate above the SC gap $\Delta_{SC}$ as their relaxation across the SC gap is suppressed by the phonon bottleneck. In SC ($T<T_c$) PuCoGa$_5$, quasiparticles accumulate in the states above the hybridization gap, similar to the photoexcitation in the normal state (Figs.~\ref{fig:elecdiag}(b),(c)). The final states during quasiparticle relaxation include both the quasiparticle states below $\Delta_{HG}$ (blue arrow in Fig.~\ref{fig:elecdiag}(c)) and the states in the SC condensate (green arrows in Fig.~\ref{fig:elecdiag}(c)). These relaxation events happen with the emission of HFPs, which creates the bottleneck and leads to long relaxation time $\tau_3$. The experimental values of $\tau_3$ at low temperatures cannot be reliably extracted from the spectra, but the observed trend in the data (Fig.~\ref{fig:ppspectra}(b)) and a good fit of T=9.2 K data obtained with fixed $\tau_3=22.9$ ps (Fig.~\ref{fig:fit}(c)) suggest that $\tau_3$ remains long, consistent with the RT model. On the other hand, the behavior of the amplitude $A_3$ deviates from the RT model prediction, which we attribute to a modification of the DOS at the Fermi level, where the SC gap opens up (Fig.~\ref{fig:elecdiag}(c)). Although a detailed description of the $A_3$ suppression cannot be constructed based solely on our data, the modification of the DOS near the Fermi level provides a plausible explanation. Thus, our data support the description of superconductivity in PuCoGa$_5$ that relies on the formation of the SC condensate from the heavy quasiparticles.

In conclusion, we measured the \eph coupling constant in PuCoGa$_5$ to be $\lambda=0.2-0.26$, which rules out phonon-mediated SC pairing. We present the first direct spectroscopic evidence for hybridization gap in the electronic DOS in the normal state and for the formation of the SC condensate from heavy quasiparticles arising from hybridization of the Pu 5$f$ electrons and the conduction electrons, both of which indicate magnetically mediated superconductivity in this material. This is consistent with a growing body of evidence\cite{curro:622, bauer:147005, chudo:083702} that suggests a $d$-wave order parameter in the transuranic superconductors (NpPd$_5$Al$_2$, PuCoGa$_5$, PuRhGa$_5$), similar to most of the other known heavy-fermion superconductors. 

This work was supported by the LDRD program at Los Alamos National Laboratory, the US DOE Office of Science, and by the Center for Integrated Nanotechnologies. E.E.M.C is supported by the Singapore Ministry of Education AcRF Tier 1 (RG41/07) and Tier 2 (ARC23/08).


\end{document}